\documentclass[11pt,11pt]{article}
\usepackage{latexsym}
\usepackage[pdftex]{graphicx}
\usepackage[]{enumerate}
\usepackage{bm}
\newcommand{\InsertFig}[5]
{\begin{figure}[#5]
      \centerline{
		\includegraphics[width=#4]{#1}
      }
      \caption{{\footnotesize  #2}}
      \label{#3}
\end{figure}}
\newcommand{\al}{\alpha}

\DeclareBoldMathCommand{\bV}{V}
\DeclareBoldMathCommand{\bF}{F}
\DeclareBoldMathCommand{\bg}{G}
\DeclareBoldMathCommand{\bv}{v}
\DeclareBoldMathCommand{\bu}{u}
\DeclareBoldMathCommand{\br}{x}
\DeclareBoldMathCommand{\bg}{g}
\DeclareBoldMathCommand{\bf}{f}
\DeclareBoldMathCommand{\bb}{b}
\DeclareBoldMathCommand{\be}{e}
\DeclareBoldMathCommand{\bs}{s}
\DeclareBoldMathCommand{\bA}{A}
\DeclareBoldMathCommand{\bB}{B}
\DeclareBoldMathCommand{\bC}{C}
\DeclareBoldMathCommand{\bD}{D}
\DeclareBoldMathCommand{\bI}{I}
\DeclareBoldMathCommand{\bM}{M}
\DeclareBoldMathCommand{\bL}{L}
\DeclareBoldMathCommand{\bN}{N}
\DeclareBoldMathCommand{\bR}{R}
\DeclareBoldMathCommand{\bS}{S}
\DeclareBoldMathCommand{\bU}{U}
\DeclareBoldMathCommand{\bk}{k}
\DeclareBoldMathCommand{\ba}{a}
\DeclareBoldMathCommand{\bn}{n}
\DeclareBoldMathCommand{\bp}{p}
\DeclareBoldMathCommand{\br}{r}

\begin{document}
\newcommand{\un}[1]{\:\mathrm{#1}}

\title{Coordinates based on a magnetic mirror field}
\author{R. D. Hazeltine}
\date{Octoberr 3, 2006
}
\maketitle
\abstract We construct a coordinate system fitting the geometry of a given, cylindrically symmetric,  magnetic field.

\section{Field line variable} \label{sec:intro}
We begin with cylindrical coordinates $(r, \theta, z)$, assuming azimuthal symmetry
\begin{equation}
\frac{\partial}{\partial \theta} = 0 \label{eq:sym}
\end{equation}
The magnetic field has the form 
\[\bm{B} = B(r,z) \bm{b}\]
where the unit vector $\bm{b}$ has components
\[
\bm{b} = (b^{r}, 0,  b^{z})\]
The field--line coordinate, $s$, measures distance along a field--line.  It is defined by
\begin{equation}
\frac{d}{ds} = b^{r}\frac{\partial}{\partial r} + b^{z} \frac{\partial}{\partial z} \label{eq:defs}
\end{equation}
This relation can be expressed as a partial differential equation for $s(r, z)$ by inserting that function into the operator on the left:  $ds/ds = 1$ or
\begin{equation}
b^{r}\frac{\partial s}{\partial r} + b^{z} \frac{\partial s}{\partial z} = 1 \label{eq:defs1}
\end{equation}

To determine the function $s(r,z)$, one first finds the field--line trajectory $\rho(\zeta;r, z)$ that intersects the point $(r, z)$, by solving
\begin{equation}
\frac{d \rho}{d\zeta} = \frac{b^{r}(\rho,\zeta)}{b^{z}(\rho,\zeta)}
\label{eq:traj}
\end{equation}
subject to the condition
\[
\rho(z;r, z) = r
\]
Next one integrates $ds = dz/b^{z}$ along the trajectory:
\begin{equation}
s(r,z) = \int_{0}^{z}\frac{d\zeta}{b^{z}(\rho(\zeta;r,z),\zeta)}
\label{eq:sint}\end{equation}
The lower integration end--point is chosen for convenience, taking note of the fact that each field line passes through the plane at $z=0$.  Thus $s$ is the distance along a field line between the point $(r,z)$ and the intersection of the line with the plane $z=0$.  
Choosing a different lower integration endpoint would add to $s$ a function $s_{0}$ that satisfies
\[
\frac{d s_{0}}{ds} = 0\]
The new $s \rightarrow s + s_{0}$ would measure the distance along a field line from some surface other than the $z =0$ plane.

It is instructive to verify that (\ref{eq:sint}) indeed solves (\ref{eq:defs1}).  The key point is that $\rho(\zeta;r,z)$, considered as a function of $r$ and $z$ for fixed $\zeta$, satisfies the homogeneous version of (\ref{eq:defs1}):
\begin{equation}
b^{r}\frac{\partial \rho}{\partial r} + b^{z} \frac{\partial \rho}{\partial z} = 0 \label{eq:key}
\end{equation}
To understand (\ref{eq:key}), note that the function $\rho(\zeta;r, z)$ is evaluated at one point on the field--line trajectory that intersects the point $(r, z)$. That is, the point $(r, z)$ fixes the particular trajectory. Now (\ref{eq:key}),
\[\frac{d \rho}{ds} = 0,\]
corresponds to incrementing $r$ and $z$ by small amounts while remaining on that same trajectory.  Since we remain on the same integral curve it is obvious that the value of $\rho$ at fixed $\zeta$ does not change.

Next we insert (\ref{eq:sint}) into the left--hand side of (\ref{eq:defs1}).  
First,
\[
b^{z}(r, z)\frac{\partial s}{\partial z} = 1 +\int_{0}^{z}d\zeta\left(\frac{\partial}{\partial \rho} \frac{1}{b^{z}}\right)b^{z}(r, z)\frac{\partial \rho}{\partial z}
\]
Similarly,
\[
b^{r}(r, z) \frac{\partial s}{\partial r} =\int_{0}^{z}d\zeta\left(\frac{\partial}{\partial \rho} \frac{1}{b^{z}}\right)b^{r}(r, z)\frac{\partial \rho}{\partial r}
\]
Adding these two expressions and using (\ref{eq:key}), we see that the integral terms cancel, leaving (\ref{eq:defs1}).

\section{Coordinates in the azimuthal plane}\label{constraints}
\subsection{Constraints}
Any vector field $\bb$ can be used to construct a function $s(r,z)$ that measures distance along the field lines; the relationship is unique, up to the arbitrariness of the $s=0$ surface.  However, when one has \emph{two} vector fields, say $\bb$ and $\ba$, the situation is more complicated, even when the vector fields are independent.

Thus suppose that we are give two vector fields, $\ba$ and $\bb$, with 
\[
\bb\cdot \nabla  = \frac{d}{ds}, \;\; \ba\cdot \nabla = \frac{d}{d\al}
\]
and that we try to construct a coordinate system $(\xi^{1}, \xi^{2}) = (\al, s)$ , replacing the system $(x^{1}, x^{2}) =(r,z)$.  A key step is to verify that the changes in $\alpha$ or $s$ between two points in space will be independent of the path linking those two points.  Thus, while $\al$ and $s$ measure  distances along the lines of their corresponding fields,  we cannot take for granted the existence of  $\alpha(r,z)$ and $s(r,z)$ as \emph{functions} on the plane. As shown in the Appendix, such functions exist only if the corresponding directional derivatives commute.   

In fact any coordinate functions must satisfy two conditions:\begin{enumerate}\item  The Jacobian must satisfy
\begin{equation}
\frac{\partial \xi^{i}}{\partial x^{j}} \frac{\partial x^{j}}{\partial \xi^{k}} = \delta^{i}_{k} \label{eq:jac}
\end{equation}
\item The directional derivatives must commute:
\begin{equation}
\left[\frac{d}{d\xi^{1}}, \frac{d}{d\xi^{2}}\right] = 0 \label{eq:com}
\end{equation} \end{enumerate}
The first condition is local, depending only on the local vector components.  The second condition, whose derivation is reviewed in the Appendix, depends upon the derivative of these components.  Neither condition is related to orthonormality; indeed we must allow our coordinates to be non-orthogonal.

Since
\begin{eqnarray}
\frac{d}{d\alpha} &=& a^{r}\frac{\partial}{\partial r} + a^{z}\frac{\partial}{\partial z} \label{eq:defa} \\
\frac{d}{ds} &=& b^{r}\frac{\partial}{\partial r} + b^{z}\frac{\partial}{\partial z} \label{eq:defb}
\end{eqnarray} 
(\ref{eq:com})  can be expressed as
\begin{equation}
(\bm{a} \cdot \nabla) b^{i} = (\bm{b}\cdot \nabla ) a^{i} \label{eq:com2}
\end{equation}

We note here that 
\begin{equation}
a^{r}  = \frac{dr}{d\alpha}, \,\,\, a^{z} = \frac{dz}{d\alpha}
\label{eq:defc}
\end{equation}
and similarly
\begin{equation}
b^{r}  = \frac{dr}{ds}, \,\,\, b^{z} = \frac{dz}{ds}
\label{eq:defb2}\end{equation}
Full derivatives, rather than partials, are used because we have yet to demonstrate that $(\alpha, s)$ are true coordinates, as the partial notation would imply.  All that is assumed here is the existence of the vector fields. 

We will derive a simple expression for the constraint (\ref{eq:com2}), which leads to an explicit (although non-unique) form for the function $\alpha({r,z})$.  But first we consider the Jacobians in more detail.

\subsection{Jacobian constraint}
 If the function $\alpha({r,z})$ exists, then it must satsify (\ref{eq:jac}):
\begin{equation} \left(\begin{array}{cc}
	\partial \alpha /\partial r & \partial \alpha /\partial z \\
	\partial s /\partial r & \partial s /\partial z 
	\end{array}\right)\left(\begin{array}{cc}
	a^{r} & b^{r} \\
	a^{z} & b^{z} 
	\end{array}\right) = \left(\begin{array}{cc}
	1 & 0 \\
	0 & 1 
	\end{array}\right)
	\label{eq:mat}\end{equation}
These equations simply express the identities
\begin{equation} \frac{d\alpha}{d\alpha} = 1,\,\,\,\frac{d\alpha}{ds} = 0, \,\,\,\frac{ds}{d\alpha} = 0, \,\,\, \frac{ds}{ds} = 1 \label{eq:loc}\end{equation}

If we suppose that the functions $s(r, z)$ and $\alpha(r, z)$ are known, then (\ref{eq:mat}) can be solved for the vector components:
\begin{eqnarray}
a^{r} = \frac{1}{\Delta} \frac{\partial s}{\partial z},& \,\,\,& a^{z} = -\frac{1}{\Delta} \frac{\partial s}{\partial r}, \label{eq:as} \\
b^{r} = -\frac{1}{\Delta} \frac{\partial \alpha}{\partial z},& \,\,\,& b^{z} = \frac{1}{\Delta} \frac{\partial \alpha}{\partial r}, \label{eq:bs}
\end{eqnarray}
Here
\begin{equation}
\Delta \equiv \frac{\partial \alpha}{\partial r}\frac{\partial s}{\partial z} - \frac{\partial \alpha}{\partial z}\frac{\partial s}{\partial r}\label{eq:det1}
\end{equation}

Since $\bm{b}$ is presumed given and $s(r, z)$ can be calculated as in Section \ref{sec:intro}, it might appear that (\ref{eq:bs}) can be solved for $\alpha(r,z)$.  But in fact the two equations of (\ref{eq:bs}) are not independent: they provide only a single independent equation for the two derivatives $\partial \alpha/\partial r$ and $\partial \alpha/\partial z$.
To see this, we substitute from (\ref{eq:det1}) to write (\ref{eq:bs}) as
\begin{eqnarray}
b^{r}\frac{\partial s}{\partial z} \frac{\partial \alpha}{\partial r} + \left(1 - b^{r}\frac{\partial s}{\partial r}\right)\frac{\partial \alpha}{\partial z} = 0 \label{eq:la1} \\
\left(1 - b^{z}\frac{\partial s}{\partial z}\right)\frac{\partial \alpha}{\partial r}+ b^{z}\frac{\partial s}{\partial r} \frac{\partial \alpha}{\partial z}  = 0
\label{eq:la2}
\end{eqnarray}
This linear, homogeneous pair is soluble only if the determinant
\[
\Delta^{\prime} \equiv b^{r}b^{z}\frac{\partial s}{\partial r}\frac{\partial s}{\partial z} - \left(1 - b^{z}\frac{\partial s}{\partial z}\right)\left(1 - b^{r}\frac{\partial s}{\partial r}\right)\]
vanishes.  Indeed it is easily seen that
\begin{equation}
\Delta^{\prime} = \bm{b} \cdot \nabla s - 1 = 0 \label{eq:detp}
\end{equation}

In summary, the two equations in  (\ref{eq:bs}) are consistent but not independent.  (Both express the relation $\bm{b}\cdot \nabla \alpha = 0$.)  They cannot by themselves determine $\alpha(r, z)$.

\subsection{Commutation constraint}
We express (\ref{eq:com2}) as $\bm{X} = 0$, where
\[
\bm{X} \equiv (\bm{a} \cdot \nabla) \bm{b} - (\bm{b}\cdot \nabla ) \bm{a}  
\]
Let us eliminate $\bm{a}$ using (\ref{eq:as}).  Beginning with the $r$--component we have
\begin{equation}
X_{r} = \frac{1}{\Delta}\left(\frac{\partial s}{\partial z}\frac{\partial b^{r}}{\partial r}  - \frac{\partial s}{\partial r}\frac{\partial b^{r}}{\partial z}  \right) - b^{r}\frac{\partial}{\partial r}\left(\frac{1}{\Delta}\frac{\partial s}{\partial z}\right) -b^{z}\frac{\partial}{\partial z}\left(\frac{1}{\Delta}\frac{\partial s}{\partial z}\right) \label{eq:xr1}
\end{equation}
We next apply the constraint (\ref{eq:defs1}), as well as the $z$--derivative of (\ref{eq:defs1}), which implies
\[
b^{r}\frac{\partial^{2}s}{\partial r \partial z} + b^{z}\frac{\partial^{2}s}{\partial z^{2}} = -\frac{\partial s}{\partial r}\frac{\partial b^{r}}{\partial z} -\frac{\partial s}{\partial z}\frac{\partial b^{z}}{\partial z}
\]
The result is
\[
X_{r} = \frac{1}{\Delta}\frac{\partial s}{\partial z}\left(\frac{\partial b^{r}}{\partial r} +\frac{\partial b^{z}}{\partial z} + \bm{b}\cdot \nabla \log \Delta \right)\]
But 
\begin{equation}
\nabla \cdot \bm{b} = \frac{\partial b^{r}}{\partial r} +\frac{\partial b^{z}}{\partial z} + \frac{b^{r}}{r} = - \bm{b}\cdot \nabla \log B \label{eq:divb} \end{equation}
so we have
\begin{equation}
X_{r} = \frac{1}{\Delta}\frac{\partial s}{\partial z} \left[\bm{b}\cdot \nabla \log \left(\frac{\Delta}{Br}\right)\right] \label{eq:xrfin}
\end{equation}
An analogous calculation shows that
\begin{equation}
X_{z} = -\frac{1}{\Delta}\frac{\partial s}{\partial r} \left[\bm{b}\cdot \nabla \log \left(\frac{\Delta}{Br}\right)\right] \label{eq:xzfin}
\end{equation}

We conclude that coordinate functions $(\alpha(r,z), s(r,z))$ exist as long as  
\begin{equation}
\Delta = rB G(\alpha) \label{eq:req}
\end{equation}
where the function $G(\alpha)$ is arbitrary.  Since
\begin{equation}
\Delta = r \nabla \theta \cdot \nabla s \times \nabla \alpha
\end{equation}
we have 
\[
B = \bB \cdot \nabla s = G(\alpha) \nabla s \cdot \nabla \alpha \times \nabla \theta
\]
Here it is clear that the function $G$ simply replaces any choice of $\alpha$ by some function of that $\alpha$, so we can choose $G = 1$.  Then (\ref{eq:bs}) provides
\begin{eqnarray}
\frac{\partial \alpha}{\partial r} &=& rB^{z} \label{eq:az}\\
\frac{\partial \alpha}{\partial z}&=& -rB^{r} \label{eq:ar}
\end{eqnarray}
 
\subsection{Example}
For concreteness we consider the mirror field given by 
\begin{eqnarray*}
B^{r} &=& B_{0}\frac{kr}{2} q \sin kz\\B^{z} &=& B_{0}(1 + q \cos kz)
\end{eqnarray*}
where $B_{0}$ and $q$ are constants.
Integrating (\ref{eq:az}) we find
\[
\alpha = qB_{0} \frac{r^{2}}{2} \cos kz + R(r) \]
Here the arbitrary function $R$ is determined from (\ref{eq:ar}), which implies
\[
R'(r) = B_{0}r\]
Hence
\begin{equation}
\alpha(r,z) = B_{0}\frac{r^{2}}{2} (1 + q \cos kz) \label{eq:afin}
\end{equation}
On the other hand $s(r,z)$, given by (\ref{eq:sint}), is a much more complicated function.  From (\ref{eq:traj}) and its boundary data we find
\[
\rho(\zeta; r,z) = r \sqrt{\frac{1 + q \cos kz}{1 + q \cos k\zeta}}\]
and therefore
\[
s(r,z) = \int_{0}^{z} d\zeta \left[ 1 + \frac{r^{2}}{4}(1 + \cos kz)\frac{k^{2}q^{2}\sin^{2}k\zeta}{(1 + q \cos k\zeta)^{3}}\right]^{1/2}\]

\section{Some consequences}
\subsection{Clebsch representations}
For an arbitrary function $f(\alpha, \theta, s)$ consider the quantity
\[
\nabla \alpha \times \nabla \theta \cdot \nabla f = (\nabla \alpha \times \nabla \theta) \cdot \nabla s \frac{\partial f}{\partial s} \]
But 
\[(\nabla \alpha \times \nabla \theta) \cdot \nabla s = \nabla \theta \cdot (\nabla s \times \nabla \alpha) = \Delta /r
\]
It follows from (\ref{eq:req}) that
\[\nabla \alpha \times \nabla \theta \cdot \nabla f = B \frac{\partial f}{\partial s} = \bm{B}\cdot \nabla f
\]
and, since $f$ is arbitrary, we can conclude that 
\begin{equation}
\bm{B} = \nabla \alpha \times \nabla \theta \label{eq:fluxrep}
\end{equation}
This relation, often used in the mirror literature, is sometimes called the flux representation of $\bm{B}$, since $\alpha$ measures magnetic flux.   The flux representation can be considered to be an alternative, ``Clebsch'' expression for the components $b^{i}$.  The corresponding expression for the components $a^{i}$ is
found in the same way and we have, in summary,
\begin{eqnarray}
B\bm{a} &=& \nabla \theta \times \nabla s \label{eq:ac}\\B\bm{b} &=& \nabla \alpha \times \nabla \theta \label{eq:bc}
\end{eqnarray}
\subsection{Orthonormality}
We can use the representation (\ref{eq:bc}) to ask whether our new coordinates are orthogonal.  The defining equations,
\begin{equation}
\bm{b} \cdot \nabla \alpha = 0 = \bm{a} \cdot \nabla s\label{eq:dot}
\end{equation}
seem to suggest orthogonality, but in fact this property requires $\bm{a}\cdot\bm{b} = 0$, which does not hold in general.  Consider, from (\ref{eq:ac})--(\ref{eq:bc}),
\[
B^{2} \bm{a}\cdot\bm{b} = (\nabla \theta \times \nabla s)\cdot (\nabla \alpha \times \nabla \theta ) \]
Since axisymmetry allows us to take $\nabla \alpha \cdot \nabla \theta = 0$, we find
\begin{equation}
r^{2} B^{2} \bm{a}\cdot \bm{b} = -\nabla \alpha \cdot \nabla s \label{eq:orth}
\end{equation}
The condition  (\ref{eq:orth}) differs essentially from (\ref{eq:dot}), because $\nabla \times \bm{b} \neq 0$ implies
\begin{equation}
\bm{b} \neq \nabla s \label{eq:bds}
\end{equation}
despite the fact that
\[
\bm{b}\cdot \nabla s = \bm{b} \cdot \bm{b} = 1\]
Thus $\nabla s = \bm{b} +\bm{e}$ with $\bm{b} \cdot \bm{e} = 0$ and neither side of (\ref{eq:orth}) can be presumed to vanish.

Geometrically the point is even simpler: field lines need not intersect a constant--$s$ surface at normal incidence.  After all, the surface on which $s$ is constant is determined by the entire ``history'' of the field line since it left the $z=0$ surface, while normal incidence is determined locally. Since this argument pertains even in the exceptional case $\nabla \times \bm{b} = 0$, we expect orthogonality of our coordinates to be rare.
\subsection{Perpendicular gradient}
The main point of field--line coordinates is the simple expression they provide for the parallel gradient,
\[
\bm{b}\cdot \nabla f = \frac{\partial f}{\partial s}\]
for  any function $f(\alpha, \theta, s)$.  Here we consider the form of the perpendicular gradient.

It is easy to compute, 
\[
\nabla_{\perp}f = \frac{\partial f}{\partial \alpha}\nabla \alpha + \frac{\partial f}{\partial \theta}\nabla \theta + \frac{\partial f}{\partial s}(\nabla s - \bm{b})\]
More useful are the components of $\bm{b} \times \nabla f$.  From the identities
\begin{eqnarray*}
\bm{b}\cdot \nabla \alpha \times \nabla \theta &=& B, \\
\bm{b}\cdot \nabla \alpha \times \nabla s &=& 0,\\
\bm{b}\cdot \nabla \theta \times \nabla s &=& B \bm{a}\cdot \bm{b}
\end{eqnarray*}
we find
\begin{eqnarray}
\nabla \alpha \cdot \bm{b } \times \nabla f& = &-B \frac{\partial f}{\partial \theta} \label{eq:ag} \\
\nabla \theta \cdot \bm{b } \times \nabla f& = &B \left(\frac{\partial f}{\partial \alpha}  - \bm{a}\cdot \bm{b} \frac{\partial f}{\partial s}\right)\label{eq:tg} \\
\nabla s \cdot \bm{b } \times \nabla f& = & B  \bm{a}\cdot \bm{b} \frac{\partial f}{\partial \theta}\label{eq:tg2}
\end{eqnarray}
\appendix
\section{Coordinate constraint}
\InsertFig{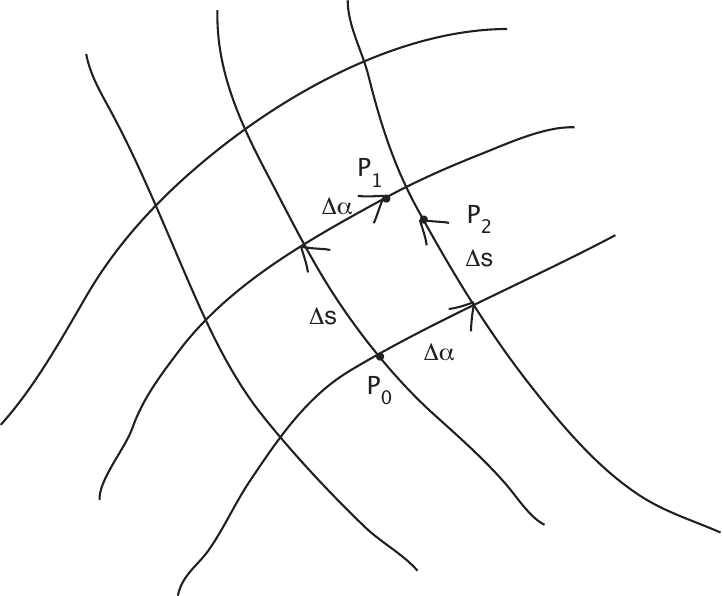}{Two paths, each beginning at $P_{0}$ and following the vector fields $\bb$ and $\ba$ in opposite order.}{fig:label}{.6\textwidth}{ht}  Here we review the derivation\cite{schutz} of the commutator constraint used in Section \ref{constraints}.  

The figure shows lines of the two vector fields $\ba$ and $\bb$. Suppose that at the point $P_{0}$ the associated field-line distances are $(\al, s)$. Beginning at $P_{0}$, we first move a distance $\Delta s$ along $\ba$, and then a distance $\Delta \al$ along $\bb$, thus ending at the point labeled $P_{1}$.  Had we performed these displacements in the opposite order, we would in general end at a distinct point $P_{2}$, also shown.  It is clear that if $\al$ and $s$ are to form coordinates in the plane, the two endpoints must be the same, $P_{1} = P_{2}$.  Indeed, both points have coordinates that differ by $(\al + \Delta \al,s+ \Delta s)$.

Now, since displacements are generated by directional derivatives, the point $P_{1}$ corresponds to 
\[
P_{1} = e^{\Delta s\, d/ds}e^{\Delta \al\, d/d\al}P_{0}
\]
Similarly
\[
P_{2} = e^{\Delta \al \,d/d\al}e^{\Delta s \,d/ds}P_{0}
\]
Expanding the exponentials and keeping terms up to second order in the displacements, we see that the two points coincide only if 
\[
[d/d\al, d/ds] = 0
\]

\bibliographystyle{pf}
\bibliography{masterbib}

\begin{thebibliography}{1}

\bibitem{schutz}
B.~Schutz,
\newblock {\em Geometrical methods of mathematical physics},
\newblock Cambridge University Press, Cambridge, 1980.

\end{thebibliography}

\end{document}